# Comparative Analysis of TELEMAC-2D Models on Agricultural Flood Damage Estimates

Théo GARIN[1,2,3], Ludovic CASSAN[1], Sophie RICCI[1], Malak SADKI[1], Quentin BONASSIES[1],
Raquel RODRIGUEZ-SUQUET[2], Sébastien LE CORRE[3]
garin@cerfacs.fr / theo.garin@cnes.fr
[1] CERFACS/CECI (CNRS UMR 5318), 42 Av. Gaspard Coriolis, 31100 Toulouse, France
[2] Centre National d'Études Spatiales (CNES), 18 Av. Edouard Belin, 31400 Toulouse, France
[3] Université Toulouse Jean Jaurès (UT2J), 5 All. Antonio Machado, 31058 Toulouse, France

*Abstract* — Direct economic impacts of flooding are essential for flood mitigation policies, and are based upon multiple tools, among which 2D hydrodynamic models. These models are dependent on multiple parameters and data, such as topography, and can yield considerably different results depending on their values, ultimately changing the resultant damage estimation. To help in understanding some of the underlying drivers of this variation, this conference paper compares damage estimations issued by two different hydrodynamic models of the Garonne River, near Marmande, in France. The influence of a topo-bathymetry projection method, which aims at reducing some interpolation incoherencies, was also studied. The general conclusion is that these two factors, i.e., a change in the model parameters and the topo-bathymetry projection method, do matter but their influence on the total cost estimation is highly dependent on the description of a few high-vulnerability agricultural fields. Indeed, the latter often account for the majority of the total damage even if occupying a fraction of the total area. This paper then advocates for a more vulnerability-based modelling approach, with a focus on the few high-vulnerability areas.

I. INTRODUCTION

Floods are among the most devastating natural disasters, causing thousands of deaths, and billions of dollars in economic losses worldwide every year [1]. More than 90% of affected people live in Asia [1], where an increase in flood occurrence was observed, and is expected to continue as such, due to multiple coupled parameters, among which anthropogenic climate change [2]. In Europe, while there is no global trend, floods in countries such as the UK, France and Germany are predicted to grow in occurrences, while diminishing in South and Eastern Europe, due to changes in precipitation [3]. Even if flood mortality per event has slightly decreased in recent years, due to advancements in research and mitigation policies, the global death toll still rises [4]. Moreover, floods tend to affect poor and marginalized people the most, which tend to represent a higher and higher part of the population due to a rise in global wealth inequality [5]. In conclusion, floods are predicted to be more numerous worldwide, their overall cost and mortality are expected to grow, and should continue to be the focus of research and attenuation policies [6, 7].

The first step in mitigating flooding events is a thorough understanding of their direct and indirect consequences. Indeed, precise comprehension of such disasters' impacts is necessary to correctly assess the effectiveness of protection measures, socio-economic policies, or land use planning whose objective are to build flood resilience [8]. Recent years saw many international efforts led in this direction. They are, among others, the International Charter Space and Major Disasters [9], the Copernicus Emergency Management Service [10], or studies conducted by the Joint Research Center of the European Commission [11]. In France such efforts are also carried out by projects such as Agirisk [12], Littoscope [13] or Flaude [14].

To assess the damages caused by one specific flood event, water hazard variables such as flood water depth, speed or residence time, has to be cross-referenced with exposure (land cover) thanks to vulnerability (or damage) models. Historically, the latter tended to be focused on direct economic impact to buildings and contents, but an increasing number of studies try to encompass a broader range of effects, such as indirect economic consequences and social impacts [8, 1, 7]. Even if direct economic damage models are well known and used, the precise estimation of damages is often still difficult and can lead to ill-adapted recovery and adaptation plans if done improperly [15].

To help in building precision when estimating direct economic impacts, it is possible to assess how water hazard modelling methods impact the resulting direct economic loss assessment. Indeed, many methods are available to characterize water hazards, but they all differ in spatiotemporal resolution, computations costs, or forecasting availability. These differences lead to different use cases, some methods are more suited for emergency assessment, while others, more costly, are more tailored for *a posteriori* estimates. But they also give different damage assessments, some more close to reality than others. Among these methods, observations regroup data from *In-Situ* (IS) measuring stations and Remote Sensing (RS) from satellites, planes or drones. IS stations can gather a wide range of flood hazard information such as water depth or speed; they offer a high temporal resolution with good measurement accuracy, but lack spatial resolution. On the other hand, RS methods can generally only assess whether there is water or not, with a limited temporal resolution, but can cover a large extent with good spatial resolution. Moreover, only using observation data does not allow for forecasting.

To fill the need of forecasting and high spatiotemporal resolution, equation-based hydrodynamic numerical models are available with solvers such as HEC-RAS [16] or openTELEMAC suite [17]. The latter contains TELEMAC-2D (T2D) that solves the Saint-Venant or Shallow Water Equations (SWE) on an unstructured 2D mesh, and is used in this study. It allows accurate simulation of water characteristics at the cost of computing power, and a careful design and calibration of the model. The water characterization provided by these models can also be improved by using Data Assimilation (DA), a family of methods that aims at increasing precision by using observation data, at the cost of high computing power. Among the latter, the Ensemble Kalman Filter (EnKF), was used to improve the water simulations on French catchments such as the Gironde estuary [18] with IS data, or the Garonne River [19, 20, 21] with IS and RS data. To reduce the need for computing power, Machine and Deep Learning (ML/DL) methods are also emerging either by directly using neural networks to simulate water hazard [22] or by incorporating other types of data into the disaster management process, like crowd-sourced or gathered on social-media information [23].

While comparisons of these hazard description methods are more common, few studies [15, 24, 25] try to assess the final impact on risk and/or losses estimation. This study aims at contributing to this effort by investigating the impact of the choice of water hazard characterisation methods on the damage estimation for a past flood event (in reanalysis mode). The main idea is to compare the different direct economic damage assessments performed with simulations issued by two 2D-hydrodynamical models based on T2D, referred to here as 'low-fidelity' (LF) and 'high-fidelity' (HF). Their main differences reside in their hydraulic structure and network description, further comparison will be made in section II.B. These comparisons will be performed on two flood events from 2019 and 2021 that occurred on the *Garonne Marmandaise* catchment (see Section II.A). The LF model is a well-known and used model that showed coherent flood simulations for both events, with and without EnKF DA [21]. The HF one is more recent and shows promising results even without DA.

The remainder of the article is organised as follows: Section II presents the data and models used in this study, whereas Section III describes the methods employed to use them. The results are detailed in Section IV. Finally, conclusions and perspectives are summarised in Section V.

II. Data

A. *Spatial extent and flood events*

The study area is the *Garonne Marmandaise* catchment, situated in the southwestern region of France. It encompasses a 50-kilometres reach of the Garonne River between *Tonneins* (upstream) and *La Réole* (downstream), whose floodplains regroup a large number of agricultural fields. In Figure 1, the general area covered by the T2D models is represented in red, while the agricultural fields these models include are depicted in blue.

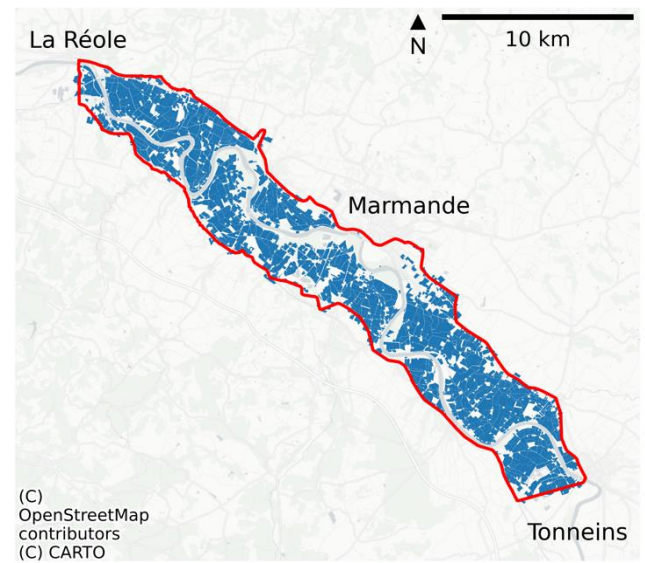

Figure 1. Models' spatial extent (red) and agricultural fields (blue) on the *Garonne Marmandaise* catchment between *Tonneins* and *La Réole*.

The floods on this catchment are mainly overflowing river events, that are – generally – not flash floods or maritime submersions. *La Réole* is the limit of the Atlantic Ocean's maritime influence. Two of these flood events, from the winters of 2019 and 2021 are simulated here. The 2019 event spanned across the last two weeks of December of the year, while the flood peak in 2021 happened in early February of the year. Hydrographs at *Tonneis* for these flood events can be found in Figure 2. To try to mitigate these events, a system of dykes and weirs was constructed incrementally to safeguard floodplains from inundation since the end of the XIX[th] century. This area also provides a data-rich environment to study floods, such as previously mentioned available T2D models, or in-situ water data which are available through the *VigiCrue* network [26], that operates observation stations at *Tonneins*, *Marmande*, and *La Réole*, providing water-level data points at 15-minute intervals. T2D models are the heart of this study, and will be presented in the next paragraph.

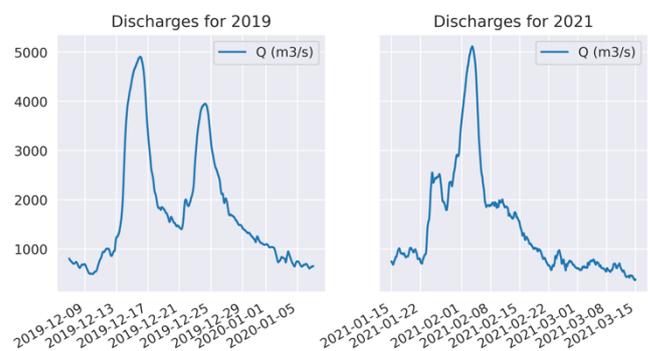

Figure 2. Hydrographs at *Tonneins* for the two flood events.

*B. T2D models, simulations and topo-bathmetry projection*

As introduced in section I, water hazard simulations for these flood events were obtained on the considered extent using two T2D-based models: an original one already used in several studies [19, 20, 21] and another one with a more refined mesh, here named LF and HF respectively. The HF mesh is based on an exhaustive detection of hydraulic structure such as dykes and roadways. These linear structures are meshed as channels, using software BlueKenue [27], by considering at least 5 points in the transverse direction: 2 at the base of structure and 3 at the top as recommended by T2D's user manual [28]. The main drainage network watercourses are also represented by smaller mesh size and use of constraint lines in the mesh building step. The influence of a better description of dykes on the flooded extent is illustrated on Figure 3, by comparing simulated water elevation at the same date (here 14th of December 2019 at noon) with the two models.

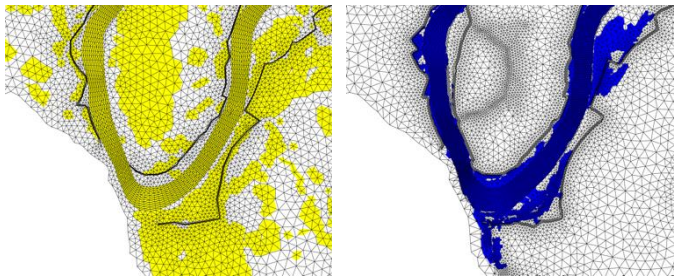

Figure 3. LF (resp. HF) simulated extent (color: wet, white: dry) and mesh on the left (resp. right) at 2019-12-14 12:00 near Meilhan-sur-Garonne.

The new and refined HF model uses the *Référentiel à Grande Échelle - ALTImétrie* (RGE ALTI) [29], as the mesh nodes elevation provider, which is a high-resolution Digital Terrain Model (DTM) (1m grid) provided by the *Institut national de l'information géographique et forestière* (IGN), a leading French public institution in the geographic field. The bathymetry in the channel mesh comes from an IS survey performed in 2023. A total of 26 cross sections are used and the interpolation between cross sections is linear in the streamwise direction, and was computed using the RAS Mapper from HEC-RAS [30]. Finally, some structures like culverts are modelled by lowering the elevation of nodes; that is, not by using the culvert object of T2D, but rather by carving the topo-bathymetry directly.

On areas such as floodplains where the mesh size is higher, regular interpolation can lead to some non-physical behaviour such as water elevation lower than the actual terrain elevation. Figure 4 illustrates this by showing that if the resolution is too low and if the terrain in between the nodes is not monotonous, it could not rightfully represent the simulated water hazard in the mesh. This situation is likely to exist in a relatively flat floodplain where buildings are set on artificially raised areas. The resultant damage estimation could be over or underestimated due to these interpolation artefacts. A way to account for this, is to project the simulated Water Surface Elevation (WSE) on a high-resolution DTM. This method is named here topo-bathmetry projection, or simply reprojection. The corrected water depth will then be the difference between the WSE and the DTM, while water velocity will be set to 0 if the DTM is higher or equal than the WSE, as indeed there cannot be velocity without water at all. The DTM used here for 'reprojecting' WSE simulated with LF and HF models, is also the RGE ALTI at 1m resolution.

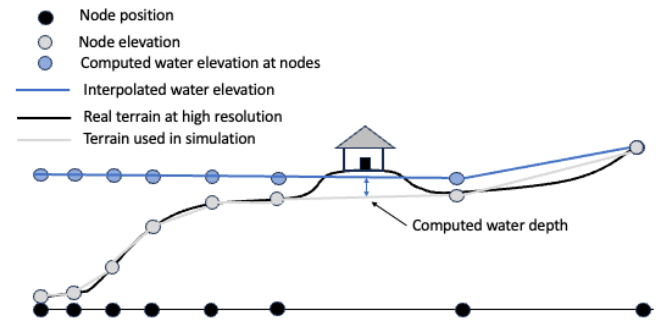

Figure 4. Illustration of interpolation incoherences for a large mesh size

This study compares a total of 8 simulations, as seen in Table 1, where the water hazard obtained with LF and HF models can be used directly to estimate the damages, or after applying a reprojection of WSE on the high-resolution DTM.

Table I Simulations' inventory

| Name | Simulation parameters | | |
|---|---|---|---|
| | *Model* | *Event* | *Reprojection* |
| LF2019 | LF | 2019 | No |
| LF2019-r | LF | 2019 | Yes |
| HF2019 | HF | 2019 | No |
| HF2019-r | HF | 2019 | Yes |
| LF2021 | LF | 2021 | No |
| LF2021-r | LF | 2021 | Yes |
| HF2021 | HF | 2021 | No |
| HF2021-r | HF | 2021 | Yes |

*C. Land cover*

The *Registre Parcellaire Graphique* (RPG) [31] is a French land cover database whose focus is the agricultural fields. It is provided yearly by the IGN. The RPG stores the agricultural fields as polygons within a shapefile (.shp) or geopackage (.gpkg), which have 3 main informations each: a unique ID allowing for easy identification of the field, a code referring to the type of culture, and the surface in hectares (ha). When direct economic impacts of simulations for the 2019 (resp. 2021) event were computed, the 2019 (resp. 2021) version of this database was used.

*D. Damage functions*

The most common damage models when trying to assess the direct economic impact of flooding are damage functions. The latter allow the conversion of water related variables such as water depth, water velocity or residence time (flood duration), to monetary costs. While very easy to use, they require a lot of expertise and work to be constructed. In France, they are built at the G-EAU laboratory (UMR INRAE-CNRS) via the R library 'floodam' [32] and distributed by the Ministry of Ecology [33]. They are empirical abacuses and not functions defined by a precise mathematical expression. These damage functions are tailored for cross referencing reference French databases provided by the IGN or the *Institut National des Statistiques et des Etudes Economiques* (INSEE), and are only relevant in France. While multiple types of these functions are available for different use cases, the ones used in this study are adapted for long term cost benefit analysis of hydraulic facilities, as they are used by local administrations during Cost-Benefit and Multi-Criteria Analyses (CBA/MCA). The analyses are performed by French territorial collectivities each time a flood prevention measure is taken: if the measure costs at least 2M€ a CBA must be conducted, and a more time-consuming MCA analysis must be made if the cost is above 5M€ [33]. Most of the time, these analyses are conducted using simulated water hazards from T2D by consultancy firms such as ARTELIA. For this reason, we wanted to use these functions to check the impact of T2D modelling methods.

For most types of land cover, the input water hazard variables are discrete water depth and residence time, aggregated on the whole flood event. In general, the maximum value of the variables during the event is kept. Maximum water levels are needed in the form of intervals: as long as the simulated continuous water level falls within the interval, the damage does not change. On the other hand, the maximum residence time has only 4 possible values: 'short', 'medium', 'long' and 'very long'. For agriculture, seasons ('winter', 'spring', 'autumn', 'summer') and discrete maximum water speed are also needed as input, the latter as 'low', 'medium' or 'high'.

The output damage is expressed in €/ha with 2016 euros; which will be written €2016/ha in this article. To have the full damage in €2016, the output damage has to be multiplied by the affected surface which can be easily gathered from the RPG database. In this study we do not update the cost estimation to the current year, to take into account factors such as inflation, but this can be easily done by following the damage functions' user guidelines, using data from the INSEE.

III. METHODS

To estimate the direct economic loss per field, we need only one value of each damage function input variable per field. For this purpose, a Monte-Carlo method is used: First, 100 points are sampled at random within each Polygon of the RPG that represent a field included in the simulation extent. A simple improvement has not been implemented here: the number of points can be adapted to the size of the polygon, the larger the polygon, the more points. Then, hydraulic variables needed as input for the damage functions, are interpolated on each of these points. They are finally aggregated to only have one value per polygon, here the mean. In order to speed up computations, the reprojection described in section II.B was only computed on the sampled Monte-Carlo points, and not every grid cell of the 1m-DTM.

To convert the 'continuous' values extracted from the simulation to the discrete ones needed by the damage functions, a per-event uniform quantile approach was used. For example, the maximum water residence time needs to fit into 4 qualitative categories that are 'short', 'medium', 'long', and 'very long'. 'Short' is then defined as anything between 0 and the 25% quantile, 'Medium' as anything between the 25% and 50% quantile and so on. Quantiles were computed on the variables' distribution spanning across the considered event only. This aggregation on one event rather than on the whole possible values taken by the variables can lead to overestimation of the hazard. For example, the maximum water speed of the event will always be characterized here as 'high', even if it is relatively low compared to other events. We do not think this would matter here, since 2019 and 2021 flood events are regarded as impactful reference events on this spatial extent.

The maximum water residence time for a flood event is here defined as the maximum consecutive amount of time where a field was flooded, and not the total amount. 'Flooded' here means that the simulated water level is above a threshold of 5cm. It was used as the water residence input, instead of simply summing the flooded days across the event. For example, if a field was flooded for one week, left to dry for another, and flooded again for 5 days, the residence time would be 7 instead of 12 days. It can lead to underestimation in the water residence time compared to summing the number of flooded days if multiple flood peaks are present. On the other hand, it allows us to take less care of the definition of the flood event, which can have a strong influence on maximum water residence times. More realistic approaches could be used, by taking into account the drying time between two flood peaks or the groundwater levels, but we opted for a simpler solution.

Some crop types such as nuts or leguminous plants did not have a corresponding damage function in the used database. For those, the damages were not computed and are therefore not represented in the results of this study. Since the LF and HF models do not strictly have the same spatial extent, we only kept the fields present in both. In the end, of a total of 3280 (resp. 3285) fields, impacts estimations are available only for 1951 (resp. 1880) of them for the 2019 (resp. 2021) flood event.

IV. RESULTS AND DISCUSSION

*A. Hydraulic differences between LF and HF models*

Figure 5 (resp. 6) shows the differences in the simulated maximum water hazard variables between the LF and HF model for the 2019 (resp. 2021) event. Indeed, as mentioned before, damage functions need an aggregated water hazard per field to be computed, here the maximum. For this reason, we want to investigate here the influence of switching from LF to HF model on these aggregated hydraulic variables, since it is mostly what will influence the computed damage. Areas in red show an underestimation compared to the HF models, while areas in blue show the opposite, an overestimation compared to the HF models.

Multiple areas are of interest here. First, the overall blue section downstream of *Tonneins* (right of the figures), on the left bank. It represents a well-known simulation artefact present in the LF model, that would overestimate the water hazard in this general area, hence the blue colour on the figures. Further downstream on the right bank, between *Tonneins* and *Marmande*, there is an overall red area regardless of the considered hazard. Here the LF model's DTM gives a topography that is a bit too high, reducing the general amount, residence time and velocity of simulated water. Downstream of Marmande, this pattern follows by showing areas where the general pattern of hydraulic variables does not change drastically between events, even if the intensity can vary. To be sure, more events should be considered, but in this configuration, the maximum water velocity map barely changes between 2019 and 2021, respectively shown in figures 5 and 6. On the other hand, water depth maps have the same overall spatial distribution between events, even if the value of the difference changes. Finally, while some areas are common between the maps, the water residence time maps seem to be much more dependent on the event. Even if more in-depth analyses could be made, the maps clearly show the influence of changing the model on the final maximum water hazards that are used as input in the damage functions.

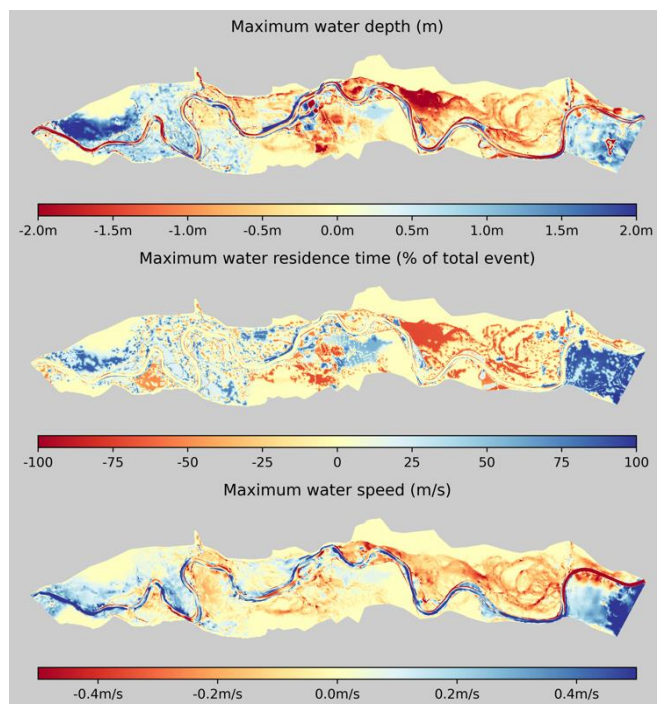

Figure 5. Differences (LF - HF) in water hazard variables for 2019 event.
Red: LF < HF | Blue: LF > HF

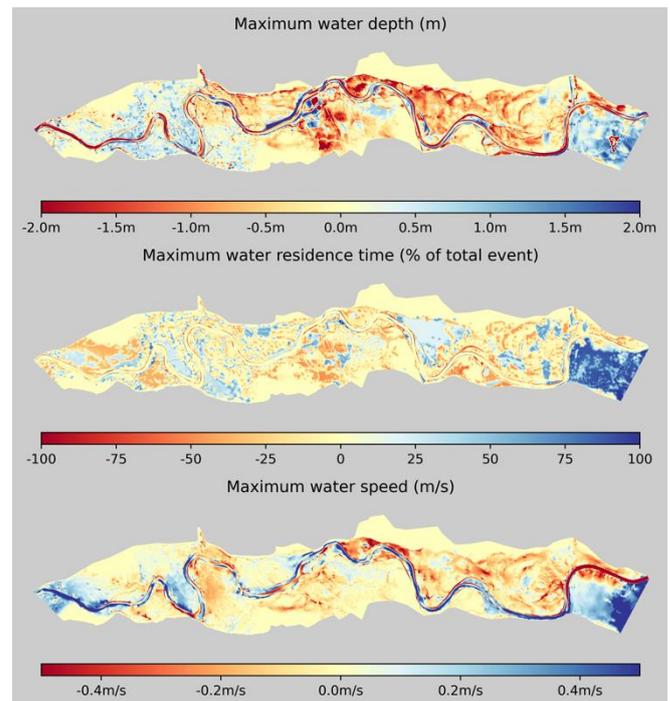

Figure 6. Differences (LF - HF) in water hazard variables for 2021 event.
Red: LF < HF | Blue: LF > HF

### B. Impacts of reprojection on water depth

The impact of reprojection of maximum water depth simulated by LF and HF models on high definition DTM, on both 2019 and 2021 events are depicted in Figure 7. The metric plotted here is the maximum water depth simulated by the models without post-treatment, minus the maximum reprojected water-depth as described in section II.B. Red colour means that the original simulation underestimates the water level compared to the reprojection, while blue shows the opposite. The figure logically shows that the importance of reprojecting on the HF model, supposedly closer to the reality, is smaller than with the LF model. Indeed, the figure shows an overall difference between simulations and their reprojected equivalent that is lower for HF derived water hazards compared to LF. There is however a narrow area near the left-bank boundary of the model where differences are significant for the HF model. This area on the left-bank corresponds to an artificial canal parallel to the Garonne that was omitted when building the model. For the latter, aside from left and right bank boundaries, a central region near *Marmande* and zones near the river minor bed seem to be the most impacted by the reprojection. For both models, it does not appear to be strongly (if at all) dependent on the considered event, even if more events should be considered to be certain.

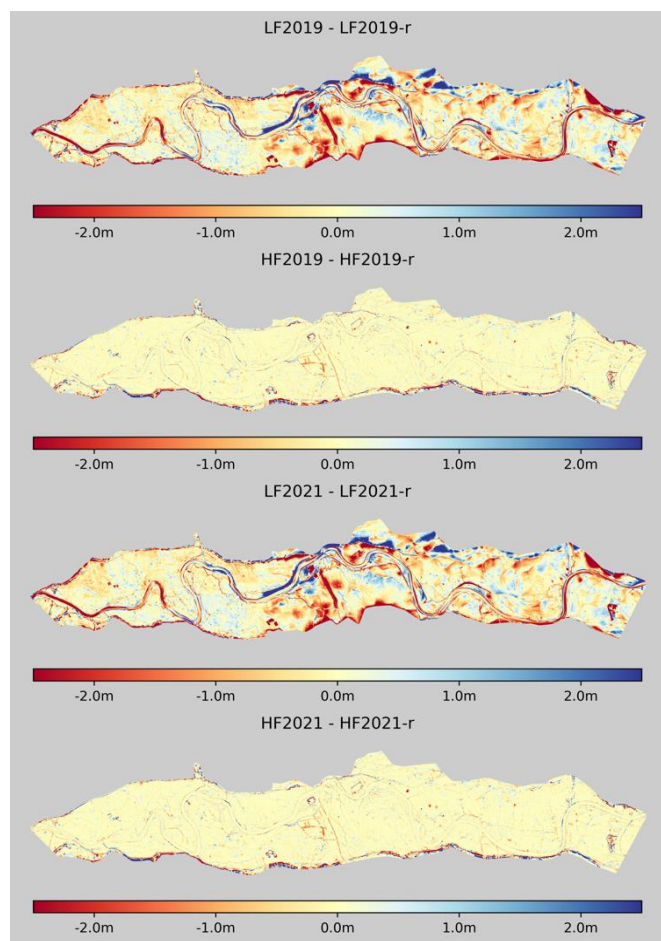

Figure 7. Differences in water depth estimations between the models and their reprojected equivalent. Red: xF < xF-r | Blue: xF > xF-r (x ∈ [L, H])

## C. Total cost and upper quantiles' influence

Table II Total direct impact estimations on agricultural fields (€2016)

| Name | Total (in M€2016 or $10^6$€2016) | | |
|---|---|---|---|
| | *Sum <95% quantile* | *Sum <99% quantile* | *All fields* |
| LF2019 | 4.18 | 6.35 | 15.7 |
| LF2019-r | 5.33 | 7.45 | 16.3 |
| HF2019 | 5.18 | 8.80 | 18.6 |
| HF2019-r | 5.51 | 8.61 | 18.5 |
| LF2021 | 4.57 | 6.29 | 9.00 |
| LF2021-r | 4.80 | 6.33 | 8.30 |
| HF2021 | 5.11 | 7.76 | 14.9 |
| HF2021-r | 5.20 | 7.24 | 14.1 |

The total damage estimation from the 8 simulations depicted in table I are presented in table II. The rightmost column gives the total estimated cost in €2016, while the second (resp. third) gives the estimation when only summing the fields under the 95% (resp. 99%) quantile of the damage distribution. This seems to allow us to draw multiple preliminary conclusions.

First, changing the model, i.e., computing damages using simulations from HF rather than LF, with or without reprojection, seems to be the most impactful driver of change of the total damage estimation. Using 2019 as an example, the LF and HF models differed by roughly 3M€, while the reprojection induced changes only accounted for 600k€ (resp. 100k€) for LF (resp. HF) simulations. Even if comparatively small, this is still a half a million euro change, showing that the reprojection technique described in II.B is not to be undermined.

Second, the cost estimations are highly dependent on the higher quantiles of the damage distribution, as indeed, the topmost 5% fields represent the majority of the total cost estimation: from half of the total for LF2021 to ⅔ for the rest. Since in France flood protection measures budget is decided based upon CBA/MCA that mainly take into account the total cost for a simulated flood event, these top 5% fields need to be analyzed more carefully.

## D. High-vulnerability crops

One way to explain this behaviour is to check the relative contribution of the experiment variables on these high-cost fields. Water hazard obviously has a role to play, as the higher it is, the higher the estimated damage. But more in-depth analyses did not show which of water depth, water velocity or residence times mattered more here in terms of producing these high-cost fields. Surface also has a role to play, as the larger the surface of the field, the more it has potential crops to be destroyed. Since in this study water hazard is aggregated per field, this can lead to unrealistic behaviour if the surface is big enough that the water hazard is radically different from one end of a field to the other. More realistic methods like subdividing a field in smaller parcels could be put in place to help in this matter. The main driver here seems to be the type of crop considered when computing damages, as come crop types seem to be more correlated to high impact fields.

Figure 8 sums the damage estimates per crop type for fields where the damage is above 20k€, which roughly correspond to the topmost 5% for each simulation. This was preferred over summing the entire distribution to better identify high vulnerability crops, and the sole influence of high cost-fields. It shows that generally, arboriculture and orchards, and some vegetables – such as cauliflower, artichokes, or broccoli, referred to here as flower-vegetables or FVs – account for a great part of the damage total. Indeed, orchards alone account for roughly 10M€ of the total damage for the 2019 event, while FVs represent between 1 and 2M€ depending on the simulation. For 2021, this trend follows, but this time with a lower total in general. Differences seen in table II regarding the total cost of the event are reflecting here. For example, in 2021 the LF simulations estimated a cumulative damage to orchards and FVs at around 3M€, while the HF gave 8 or 9M€. This 5 to 6M€ difference is reflected in the total estimate, shown in the rightmost column of table II. Some crops, such as corn, seem to be relevant here simply because of the large number of corn fields, as shown in Figure 9.

On the other hand, the figure also shows that orchards and FVs only account for a few fields, meaning that their estimated cost is mostly driven by their high flood vulnerability.

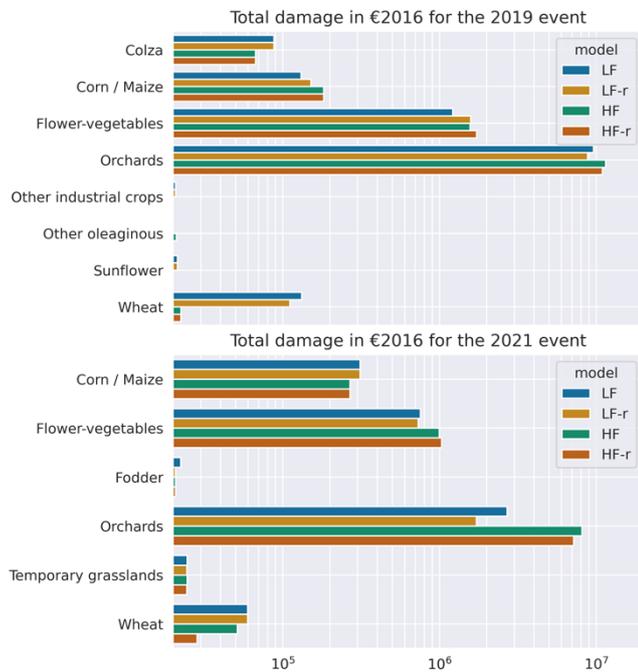

Figure 8. Total contribution to the total score of >20k€ fields per crop type

When looking at the damage functions used to make these estimates, these results are not really groundbreaking, as some crops are indeed more expensive to replace than others. For example, for a maximum water hazard in depth, velocity and residence time (which happened at least for some orchards in our simulations), the surface cost given by the damage functions reaches around 93k€/ha for orchards in the winter, 51k€/ha for grapevines, 6.6k€/ha for FVs, while only up to 1 or 2k€/ha for the other crop types.

This can be explained because permanent crops such as orchards or grapevines – the latter not appearing in Figure 8 due to the small size of the fields on this spatial extent, hence their individual impact lower than 20k€ – take years to grow and produce fruits. If they are wiped during a flood, the damage takes into account these recovery years with no production, while fast growing crops could just be harvested the next year. This also puts the estimated damages for subsequent events in perspective. If orchards were destroyed in 2019 and replanted soon after, then when the trees were flooded again in 2021, they were not at full maturity yet.

In any case, these findings seem to advocate for a more vulnerability-based modelling. Indeed, if some types of permanent crops represent more than half of the total estimate while only occupying a small proportion of the total extent, meshes and topography should be refined and really considered with care on these areas. Moreover, as the crops are permanent, the refining effort would only be a one-time operation.

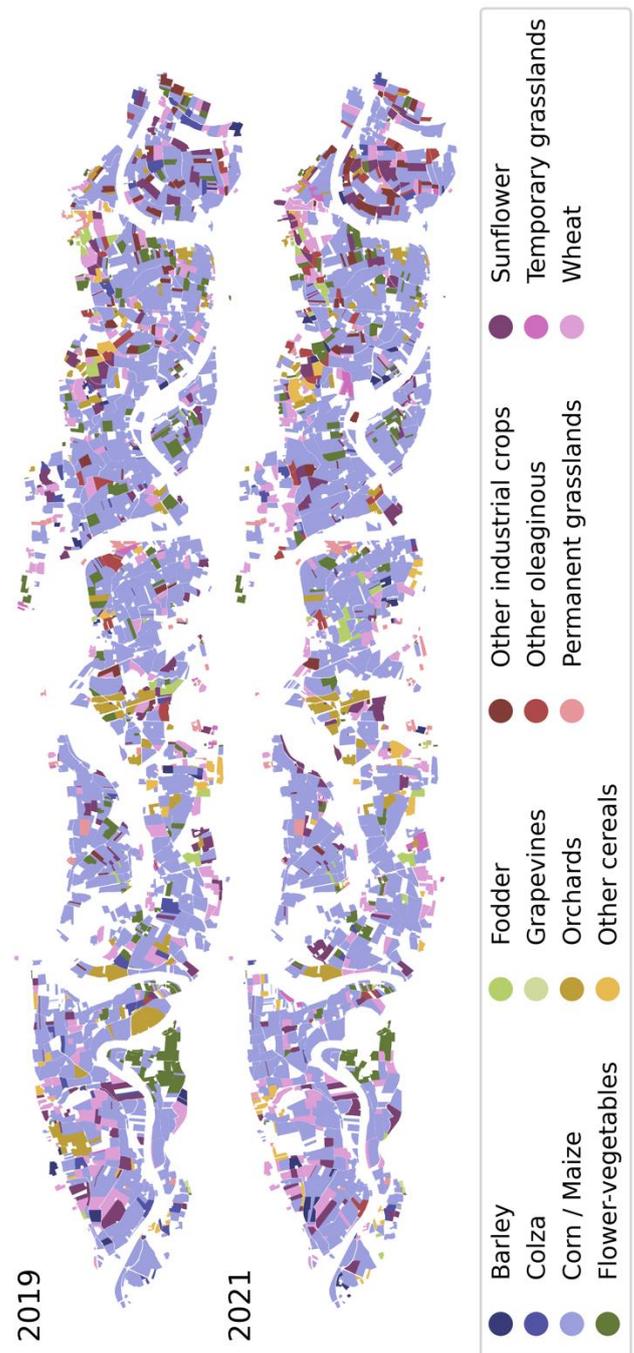

Figure 9. Crop types extracted from RPG databases for 2019 and 2021

E. *Empirical Cumulative Distribution Functions*

Another way of characterising these high cost and vulnerability fields is to build Empirical Cumulative Distribution Functions (ECDF) of the computed damages (in €2016), which are plotted in Figure 10, with a close-up on quantiles above 0.95. This ECDF shows on its horizontal axis the flood damage in €2016 and its corresponding quantile on the vertical axis.

For example, the 0.75 quantile roughly corresponds to a damage of 3k€2016, meaning that approximately 75% of agricultural fields have a damage lower than or equal to 3k€.

For all simulations, a general conclusion can be drawn from Figure 10: roughly 95% of fields account for less than 20k€ each, while some fields can reach 2M€ for the highest quantiles, often and mostly because of their crop type. The ECDFs seem to be somewhat dependent on the event, but general conclusions can still be drawn. On the two considered events, reprojection has a strong relative impact on the lower quantiles but a light relative impact on the higher quantiles. For the low quantiles on the 2019 event as an example, 30€ corresponds to the 35% quantile for LF, while only 7.5% for LF-r. For HF in this specific case, it goes from roughly 23 to almost 0%. The modelled damage for a given field can indeed only be in the tens of euros, when water hazard, surface and crop cost are low altogether.

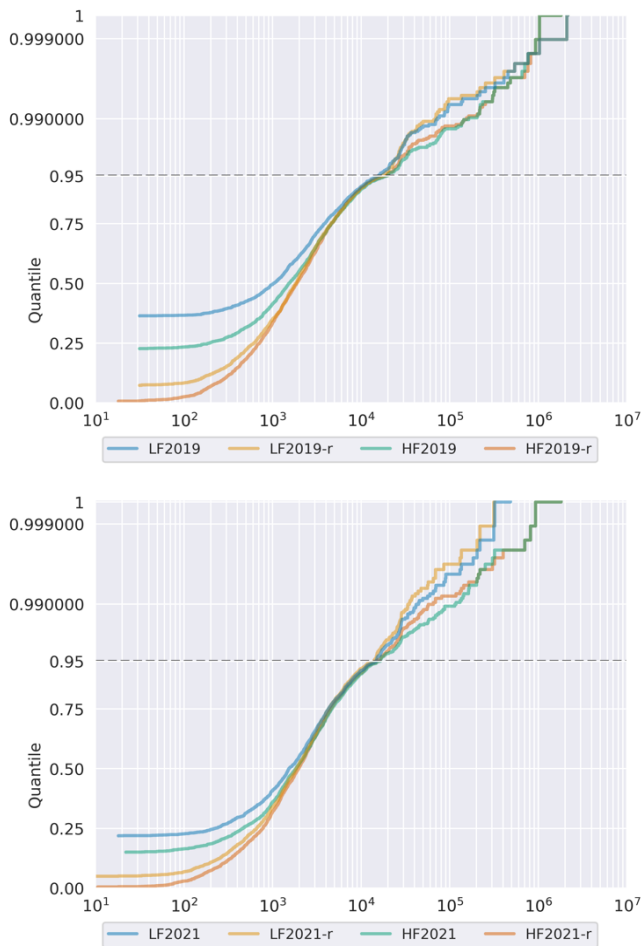

Figure 10. Empirical Cumulative Distribution Functions (ECDFs) of the direct damage (€2016) estimated by the different simulation sources.

This shift of the lower quantiles shows that damages computed with reprojected water depth have a generally higher minimum than their non reprojected equivalent. On the other hand, with the higher quantiles, reprojection seems to have less relative impact, with a slight tendency to lower the maximum damage distribution. Even if relatively small, quantile values changes can still range from $10^4$ to $10^5$€. In summary, reprojection seems to slightly increase the value of lower damages, while slightly decreasing the value of higher damages. In other words, reprojection seems to marginally reduce the global spread of estimated damages. The differences between LF and HF based simulations can also be observed in the same way. On the lower quantiles, reprojection is more impactful, but still significant between LF and HF. Indeed, by taking the same values as before, 30€ corresponds to the 35% quantile for LF while only 23% for HF, showing that the latter's lower damages seem to then be slightly increased compared to LF. HF induced damages are also increased in the upper quantiles, as the quantiles values are higher than their LF equivalent. This shift in the upper quantiles is more important when changing the model than with reprojection. The different behaviours of higher quantiles among simulations, and how it differs based on changing from LF to HF model or reprojecting can help understand how impactful these parameters may be on the total damage estimation. Indeed, higher quantiles definitely have a strong impact on total losses, of which they generally account for two-thirds. The fact that high quantile damages obtained from HF simulations, reprojected or not, are often higher than their LF equivalent, explain the fact that the total estimation is generally higher among events with HF rather than LF, as shown in table II. However, the same conclusion is not true for the reprojection technique, that tends to slightly lower the upper quantile damages, but table II does not allow us to see this reflect on the total estimation. More events should be to be considered to study the generalization of this behaviour.

*F. Spatial repartition of the damages*

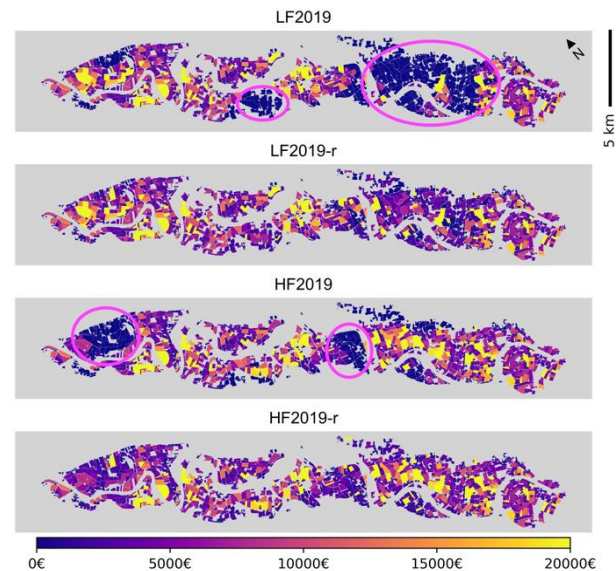

Figure 11. Direct impact estimations for the 2019 event (< 20k€)

The spatial distribution of the damages is shown in Figure 11 and 12, for the events 2019 and 2021 respectively, with their values limited to 20k€ to see global spatial trends and effects of hydraulic infrastructures more clearly.

Primary conclusions can be drawn here and will be discussed again in the following paragraphs. First, the high impact fields match the high vulnerability crops, that are orchards or FVs, mapped on Figure 10. Then, the reprojection effect on the lower quantiles described in the previous section is reflected here, where 0-cost fields (deep blue, e.g., pink ellipses on Figure 11) have often a higher cost when reprojected. While for 2019 the overall spatial distribution varies between models on some specific areas (Figure 11), they are much more uniform for the 2021 event (Figure 12). To try and see these changes more precisely, we try to assess the relative contributions of reprojection and changing the model when estimating the flood direct economic impact in the following paragraphs.

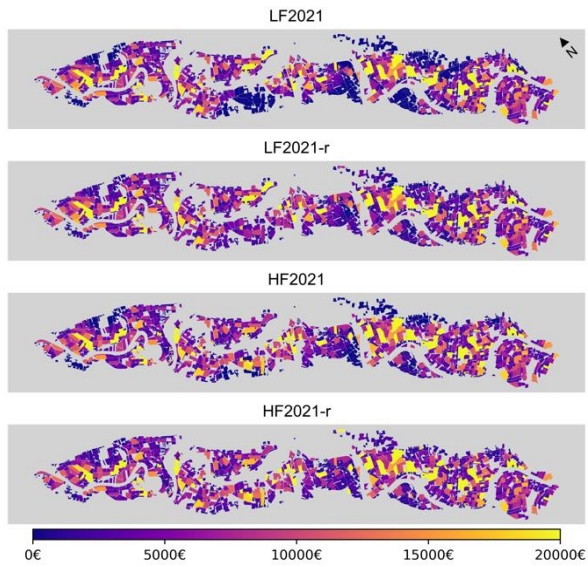

Figure 12. Direct impact estimations for the 2021 event (< 20k€)

### G. *Spatial impact of changing the model*

Figure 13 shows pairwise comparison between the LF and HF models for events 2019 and 2021. Precisely, the differences LF-HF are plotted, with their values clipped to +/- 7.5k€ – which roughly corresponds to the 95 quantile of the difference's distribution – to better understand global spatial trends. First of all, the spatial distributions roughly match the water depth differences map shown in Figures 5 and 6 respectively for 2019 and 2021. That is for 2019, an area downstream of *Tonneins* on the right bank that is underestimated by the LF model (red ellipse on the figure), and an area upstream of *La Réole* on the right bank that was overestimated by the LF model (blue ellipse). For both events, the central region on the left bank near *Marmande* is also coherent with the differences in the water height differences plotted in Figures 5 and 6 (purple ellipse). Even if water depth seems at first glance to be the main driver here, the importance of water residence time and velocity can be seen on the damage estimate, for example in the area delimited by the black ellipse on Figure 13. Indeed, despite a high water level difference as shown in Figure 6, the damage difference is non-existent, probably because the residence time's difference cancelled out the depth difference in the multivariate damage function. In the same way, Figure 14 depicts damage issued from simulations that were both reprojected. It shows a similar spatial distribution of damages even if the intensity of the differences is generally smaller. This attenuation could be explained by the topo-bathymetry reprojection induced changes depicted in Figure 7, even if a clear conclusion seems hard to draw here.

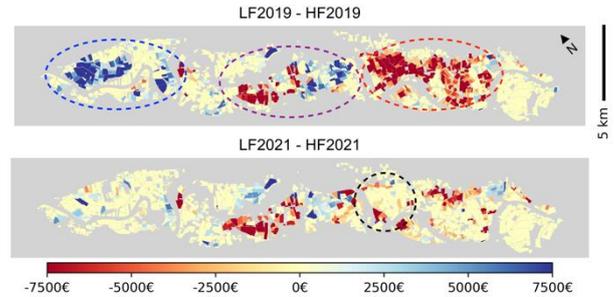

Figure 13. Differences in cost estimation between LF and HF models.
Red: LF < HF | Blue: LF > HF

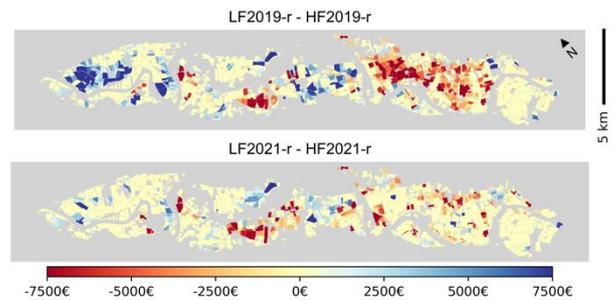

Figure 14. Differences in cost estimation between LF-r and HF-r models.
Red: LF-r < HF-r | Blue: LF-r > HF-r

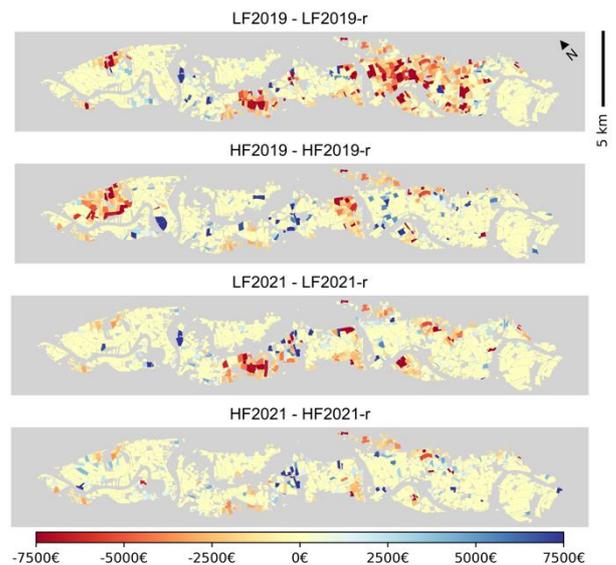

Figure 15. Differences in cost estimation between the models and their reprojected equivalent. Red: xF < xF-r | Blue: xF > xF-r (x ∈ [L, H])

*H. Spatial impact of reprojection*

The reprojection impact of the damage estimation plotted in Figure 15 seems to be quite sensitive to the used model and considered event. On the other hand, maximum water depth differences plotted in Figure 7, while showing a difference based on the model, were pretty consistent between the two events. This could be explained by the other water hazard metrics used as impact by the damage functions, as they are indeed also impacted by reprojection, as mentioned in section II.B. Overall this method seems to increase the damage of low impact fields and reduce the ones of some specific high vulnerability crops, in blue on the figure and whose crop type can be seen on Figure 9.

V. CONCLUSION

This conference paper aimed at estimating the direct economic impact of two flooding events on agricultural fields, by using water hazard simulated using two T2D based models. The main differences in these models, LF and HF, reside in their description of the topography and hydraulic infrastructures. To also help in this matter, a reprojection method was also used and its impact on the damage estimation was investigated. First of all, between the impact on the damage estimate of the reprojection technique and the model change, the latter has a stronger importance on both the total damage estimate and the spatial distribution of the damage. However, it is a costly operation to update a model – even if its influence is crucial and should be done anytime it is possible – while the reprojection method is relatively computationally cheap and can be done systematically, to make simulation results more realistic. To further investigate the influence of this topo-bathymetry projection method, it would be interesting to build a coherent T2D model that has the same spatial resolution as the RGE ALTI DTM and run the same analyses, which we guess would show a lesser influence of the reprojection method. However, such models would be impossible to run in reasonable times on current computers, as solving the SWEs for a 1m mesh on a 50km reach is way too costly for these flood events.

The other main point of this work is that given that certain crop types are more vulnerable than others, and that their share of the total cost estimation is major, more vulnerability-based modelling efforts should be conducted in order to better represent these high importance fields. Indeed, even if the differences are only small, this could still result in a drastic impact on the total damage estimate. These high vulnerability crops, such as orchards or grapevines, are often permanent so the modelling effort of refining the water hazard simulations on these fields is an investment that would pay off for multiple years.

This work also has lot of limitations. First of all, we focused on one specific area that has a given topography, climate and type of crops, and on two flood events with an overflowing regime. The data, models and methods used can also be discussed and improved. This does not necessarily allow for a generalisation of these findings, as the latter would need to consider a greater diversity of spatial extents, crop types and flood regimes.

Nonetheless, a vulnerability-based approach when it comes to hydrodynamic modelling seems reasonable, as this is ultimately what drives public funding for flood protection measures, at least in France.

A perspective of this work is to take more observation or simulation methods into account, like satellite-derived water maps, 1D or AI models, and see their impact on the water hazard and resultant damage estimations compared to the default 2D modelling technique.

ACKNOWLEDGEMENTS

This work was conducted during T. GARIN's PhD thesis, funded by the CNES and in collaboration with CECI/CERFACS and UT2J.

REFERENCES

[1] Merz, B., Blöschl, G., Vorogushyn, S., Dottori, F., Aerts, J. C., Bates, P., ... & Macdonald, E. (2021). Causes, impacts and patterns of disastrous river floods. *Nature Reviews Earth & Environment*, *2*(9), 592-609.

[2] Wang, X., Xia, J., Dong, B. *et al.* Spatiotemporal distribution of flood disasters in Asia and influencing factors in 1980–2019. *Nat Hazards* 108, 2721–2738 (2021). https://doi.org/10.1007/s11069-021-04798-3

[3] Blöschl, G., Hall, J., Viglione, A., Perdigão, R. A., Parajka, J., Merz, B., ... & Živković, N. (2019). Changing climate both increases and decreases European river floods. *Nature*, *573*(7772), 108-111.

[4] Hu, P., Zhang, Q., Shi, P., Chen, B., & Fang, J. (2018). Flood-induced mortality across the globe: Spatiotemporal pattern and influencing factors. *Science of the Total Environment*, *643*, 171-182.

[5] Lindersson, S., Raffetti, E., Rusca, M., Brandimarte, L., Mård, J., & Di Baldassarre, G. (2023). The wider the gap between rich and poor the higher the flood mortality. *Nature Sustainability*, *6*(8), 995-1005.

[6] IPCC, 2023: Summary for Policymakers. In: Climate Change 2023: Synthesis Report. Contribution of Working Groups I, II and III to the Sixth Assessment Report of the Intergovernmental Panel on Climate Change [Core Writing Team, H. Lee and J. Romero (eds.)]. IPCC, Geneva, Switzerland, pp. 1-34, doi: 10.59327/IPCC/AR6-9789291691647.001

[7] Sendai Framework for Disaster Risk Reduction: https://www.undrr.org/publication/sendai-framework-disaster-risk-reduction-2015-2030

[8] Allaire, M. (2018). Socio-economic impacts of flooding: A review of the empirical literature. *Water Security*, *3*, 18-26.

[9] International Charter Space and Major Disasters: https://disasterscharter.org/web/guest/home

[10] Copernicus EMS: https://emergency.copernicus.eu/

[11] Joint Research Center - European Commission: https://joint-research-centre.ec.europa.eu/index_en

[12] AgiRisk, Cerema: https://agirisk.cerema.fr/

[13] SCO Littoscope: https://www.spaceclimateobservatory.org/fr/littoscope

[14] SCO FLAude: https://www.spaceclimateobservatory.org/fr/flaude

[15] Sieg, T., & Thieken, A. H. (2022). Improving flood impact estimations. *Environmental Research Letters*, *17*(6), 064007.

[16] HEC-RAS: https://www.hec.usace.army.mil/software/hec-ras/

[17] OpenTelemac: https://www.opentelemac.org/

[18] Laborie, V. (2020). Quantification d'incertitudes et assimilation de données pour la modélisation hydrodynamique bidimensionnelle: application au modèle de prévision des hautes eaux de l'estuaire de la Gironde (Doctoral dissertation, Université Paris-Est).

[19] Nguyen, T. H., Ricci, S., Fatras, C., Piacentini, A., Delmotte, A., Lavergne, E., & Kettig, P. (2022). Improvement of flood extent

representation with remote sensing data and data assimilation. *IEEE Transactions on Geoscience and Remote Sensing*, *60*, 1-22.

[20] Nguyen, T. H., Ricci, S., Piacentini, A., Fatras, C., Kettig, P., Blanchet, G., ... & Baillarin, S. (2022). Dual State-Parameter Assimilation of SAR-Derived Wet Surface Ratio for Improving Fluvial Flood Reanalysis. *Water Resources Research*, *58*(11), e2022WR033155.

[21] Nguyen, T. H., Ricci, S., Piacentini, A., Simon, E., Rodriguez-Suquet, R., & Luque, S. P. (2023). Gaussian anamorphosis for ensemble kalman filter analysis of SAR-derived wet surface ratio observations. *IEEE Transactions on Geoscience and Remote Sensing*.

[22] Bentivoglio, R., Isufi, E., Jonkman, S. N., & Taormina, R. (2022). Deep learning methods for flood mapping: a review of existing applications and future research directions. *Hydrology and Earth System Sciences Discussions*, *2022*, 1-50.

[23] Wang, R. Q. (2021). Artificial intelligence for flood observation. In *Earth observation for flood applications* (pp. 295-304). Elsevier.

[24] Ernst, J., Dewals, B. J., Detrembleur, S., Archambeau, P., Erpicum, S., & Pirotton, M. (2010). Micro-scale flood risk analysis based on detailed 2D hydraulic modelling and high resolution geographic data. *Natural Hazards*, *55*, 181-209.

[25] Sieg, T., Kienzler, S., Rözer, V., Vogel, K., Rust, H., Bronstert, A., ... & Thieken, A. H. (2023). Toward an adequate level of detail in flood risk assessments. *Journal of Flood Risk Management*, *16*(3), e12889

[26] VigiCrues: https://www.vigicrues.gouv.fr/

[27] BlueKenue: https://nrc.canada.ca/index.php/fr/recherche-developpement/produits-services/logiciels-applications/blue-kenuetm-logiciel-modelisateurs-hydrauliques

[28] TELEMAC-2D user manual v7p0: https://www.opentelemac.org/downloads/MANUALS/TELEMAC-2D/telemac-2d_user_manual_en_v7p0.pdf

[29] RGE ALTI: https://geoservices.ign.fr/rgealti

[30] HEC-RAS RAS Mapper documentation : https://www.hec.usace.army.mil/confluence/rasdocs/rmum/latest

[31] RPG: https://geoservices.ign.fr/rpg

[32] Floodam: https://www.floodam.org/

[33] Ministry of Ecology damage functions: https://www.ecologie.gouv.fr/levaluation-economique-des-projets-gestion-des-risques-naturels